\documentclass{jaa}

\usepackage{natbib}
\usepackage{graphicx}	
\usepackage{url}
\usepackage{hyperref}
\usepackage{siunitx}

\newcommand\astrosat{{\it AstroSat}}

\newcommand\arcmin{{\rm~arcmin}}
\newcommand\arcsec{{\rm~arcsec}}

\newcommand\aj{AJ}

\newcommand\apj{ApJ}

\newcommand\apjs{ApJS}

\newcommand\pasp{PASP}

\newcommand\procspie{Proc.~SPIE}

\begin{document}\sloppy

\title{Calibration of \astrosat{}/UVIT Gratings and Spectral Responses}


\author{G. C. Dewangan\textsuperscript{1}}
\affilOne{\textsuperscript{1}Inter-University Centre for Astronomy and Astrophysics (IUCAA), SPPU Campus, Pune, India.\\}


\twocolumn[{

\maketitle

\corres{gulabd@iucaa.in}

\msinfo{7 November 2020}{20 December 2020}

\begin{abstract} 
\astrosat{}/UVIT carries two gratings in
the FUV channel and a single grating in the  NUV channel. These
gratings are useful for low resolution, slitless spectroscopy in the
far and near UV bands of a variety of cosmic sources such as hot
stars, interacting binaries, active galactic nuclei, etc. We present calibration of these gratings
using observations of  UV standards NGC~40 and HZ~4. We perform
wavelength and flux calibration and derive effective areas for
different grating orders. We find peak effective areas of
$\sim18.7{\rm~cm^2}$ at $2325\AA$ for the $-1$ order of NUV-Grating, 
$\sim4.5{\rm~cm^2}$ at $1390\AA$ for the $-2$ order of FUV-Grating1, and
$\sim 4.3{\rm~cm^2}$ at $1500\AA$ for the $-2$ order of FUV-Grating2.  The
FWHM spectral resolution of the FUV gratings is $\approx14.6\AA$ in the
$-2$ order. The $-1$ order of NUV grating has an FWHM resolution of
$\approx33\AA$.  
We find excellent agreement in flux
measurements between the FUV/NUV gratings and all broadband filters.
We have generated spectral response of the UVIT gratings and broadband
filters that can directly be used in the spectral fitting packages such as
{\tt XSPEC}, {\tt Sherpa} and {\tt ISIS}, thus allowing spectral analysis of UVIT data either separately or jointly with X-ray data
from \astrosat{} or other missions.

\end{abstract}

}]


\doinum{12.3456/s78910-011-012-3}
\artcitid{\#\#\#\#}
\volnum{000}
\year{0000}
\pgrange{1--}
\setcounter{page}{1}
\lp{1}

\section{Introduction}
The Ultra-Violet Imaging Telescope (UVIT; \citealt{2016SPIE.9905E..1FS,Tandon_2017,2020AJ....159..158T}) is one of the four co-aligned payloads on-board the Indian multi-wavelength space observatory \astrosat{} \citep{2006AdSpR..38.2989A,2014SPIE.9144E..1SS}. UVIT is a twin telescope system, one of them is designed to observe in the far ultra-violet band ($1300-1800\AA$) and is known as the FUV channel. The second telescope utilizes a beam splitter that separates near UV and visible light, thus forming two detection channels -- near UV (NUV; $2000-3000\AA$) and visible (VIS; $3200-5500\AA$). The three channels use identically configured intensified CMOS detector systems which only differ in having different photo-cathodes as per the wavelength band, and operate simultaneously. The beam splitter and the mechanical mounting of the telescopes cause the NUV  channel field to be inverted and rotated by $33^{\circ}$ with respect to the FUV channel field. Each channel is equipped with a set of filters that allow selection of bands with spectral coverage of $\Delta\lambda\sim 125\AA-500\AA$ (FUV), $\Delta\lambda\sim 90\AA-800\AA$ (NUV) and $\Delta\lambda\sim 400\AA-2200\AA$ (VIS).  The FUV and NUV channels provide excellent imaging capability with a point spread function (PSF) in the range  $1-1.5\arcsec$ (FWHM). The UVIT is mainly an imaging instrument with limited spectral capability. It can be used for low resolution, slitless spectroscopy in the far and near UV bands.

\begin{figure*}
\centering
\includegraphics[width=5.4cm]{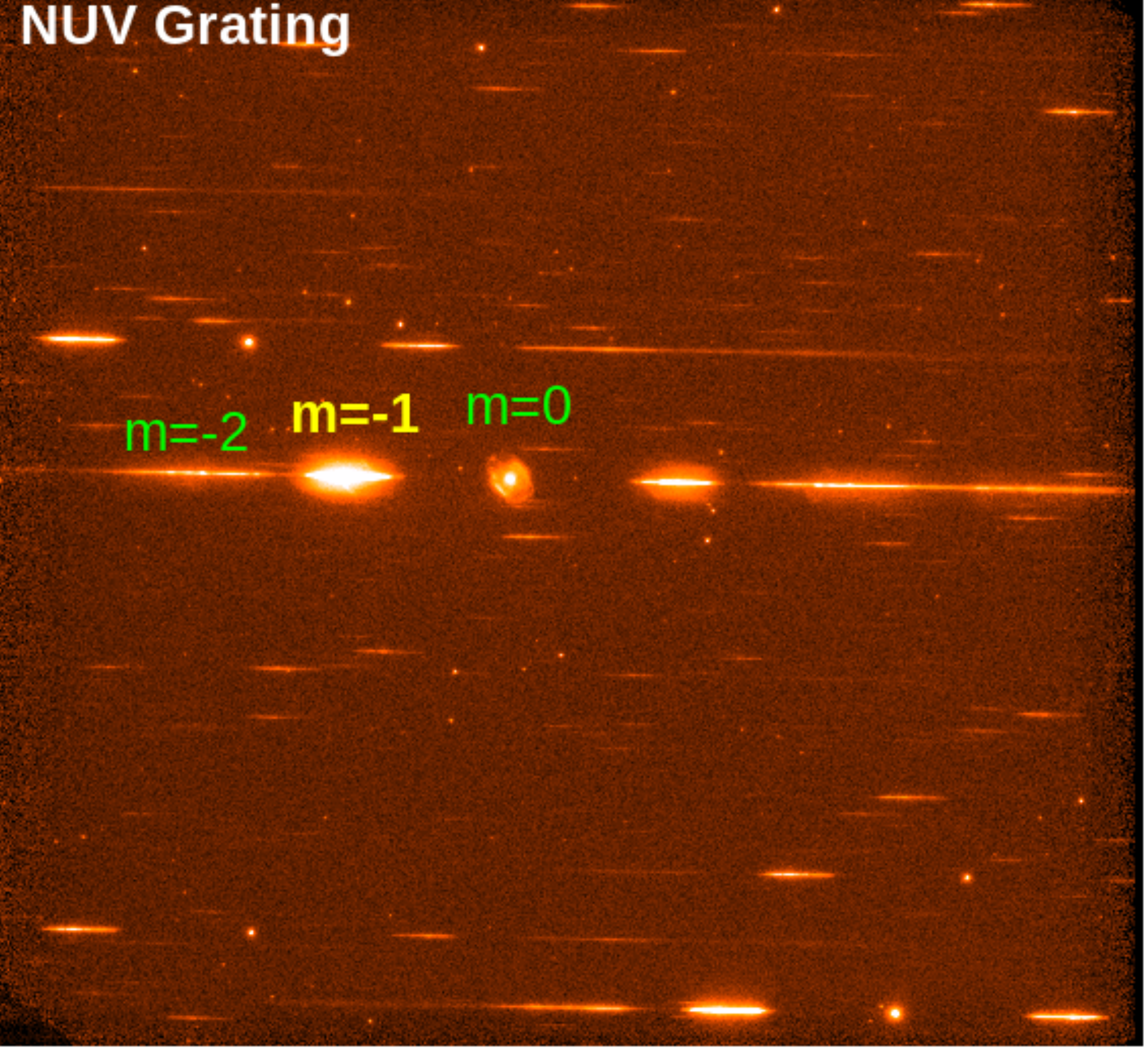}
\includegraphics[width=5.3cm]{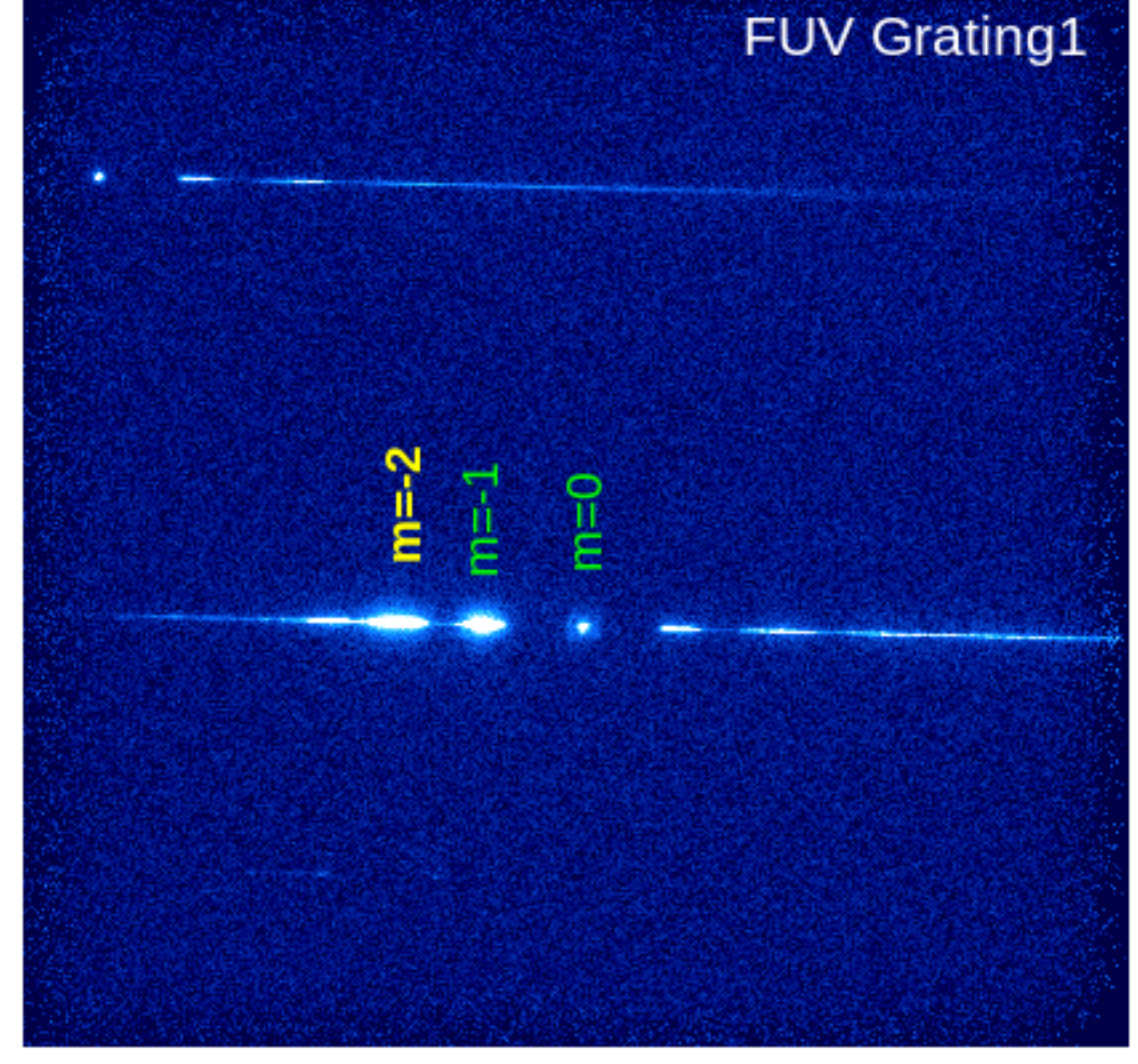}
\includegraphics[width=5cm]{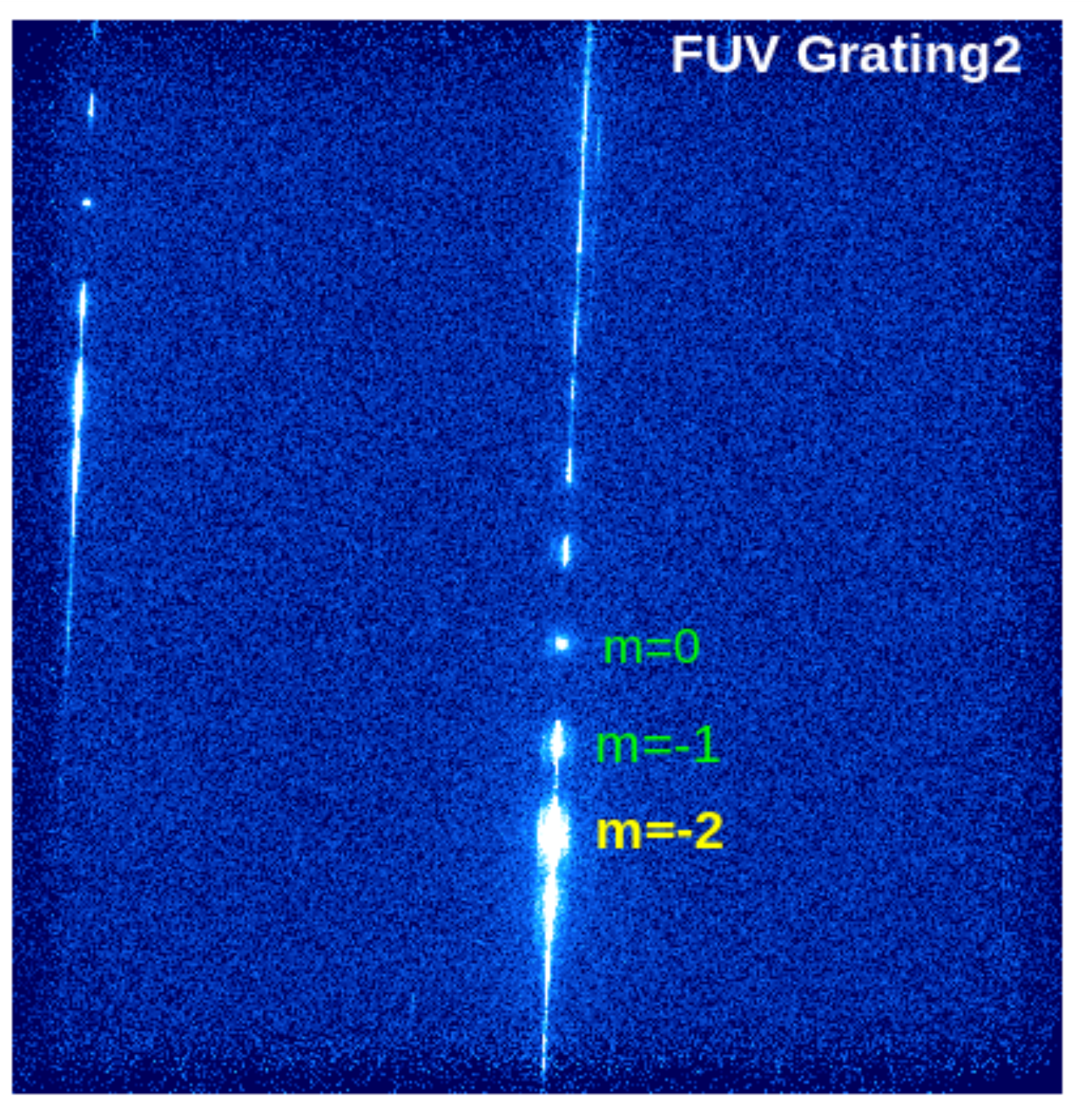}

\caption{FUV/NUV Grating images of NGC40. The negative grating orders are marked. The image sizes are $\sim9.8\arcmin$ on one side.}
\label{fig:grating_images}
\end{figure*}

There are three UVIT gratings that are ruled with $400{\rm~lines~mm^{-1}}$ on CaF2 substrates of 4.52~mm thickness. The dispersion in the detector plane caused by each grating is $12\AA~\arcsec^{-1}$ and $6\AA~\arcsec^{-1}$ in the first and second order, respectively, at  $1350\AA$. One of three gratings is mounted on the NUV channel filter wheel while the other two  gratings are mounted on the FUV  channel filter wheel such that their dispersion axes are nearly perpendicular. Such an  orthogonal arrangement of FUV gratings helps in avoiding contamination of nearby sources in the grating images. A source along the dispersion arm of the main target in one FUV grating image will cause contamination  while the dispersion arms will be well separated in the other FUV grating image due to the orthogonal arrangement. The FUV gratings and the detector are designed to maximize the efficiency in the $m=-2$ order while the NUV grating and the detector provide maximum efficiency in the $m=-1$ order. The main parameters of the gratings are listed in Table~\ref{tab:grating_par}.  More details on UVIT gratings can be found in \cite{Tandon_2017,2020AJ....159..158T}. The UVIT gratings have been described with different names, in Table~\ref{tab:grating_par} we list all the IDs  to avoid any confusion when using different resources. Here we refer the gratins as FUV-Grating1 (FUV-G1), FUV-Grating2 (FUV-G2) and NUV-Grating (NUV-G).

In this paper, we present calibration of the UVIT gratings. Some of the results derived in this paper are presented in \cite{2020AJ....159..158T}. 
Here, we present the calibration of the UVIT gratings, some results of which can be found in Tandon et al 2020.  Our updated work here  results in minor changes to the wavelength and flux calibrations, and also includes additional grating orders.
We describe the calibration method, derive additional calibration products and discuss cross-calibration between the gratings and broadband filters. The paper is organized as follows. We describe the UVIT data and the  reduction in Sec.~\ref{sec:data}, extraction of one dimensional (1d) grating spectra in Sec.~\ref{sec:extr_spec}, and wavelength calibration in Sec.~\ref{sec:wave_calib}. We derive effective areas and perform flux calibration in Sec.~\ref{sec:flux_calib}, and discuss cross-calibration between the gratings and broadband filters in Sec.~\ref{sec:cross_calib}
In Sec.~\ref{sec:spec_resp}, we derive instrumental response in the form of a product of redistribution matrix and ancilliary response that can be directly used in popular spectral fitting packages in X-ray astronomy. Finally we summarize our results in Sec.~\ref{sec:summary}.

\begin{table*}
\centering
\caption{UVIT Grating parameters}

\begin{tabular}{cccc} \hline 
Parameter & FUV-Grating1 & FUV-Grating2 & NUV-Grating \\ \hline
Filter Wheel Slot number &  4  &  6 & 4 \\ 
IDs in APPS & 4 - grating1 (FUV) & 6 - grating2 (FUV) & 4 - grating (NUV) \\
IDs in CCDLAB & FUV\_Grating1 & FUV\_Grating2 & NUV\_Grating \\ 
This paper & FUV-G1 & FUV-G2 & NUV-G \\
IDs in \cite{2020AJ....159..158T} & FUV1 & FUV2 & NUV \\
IDs in \cite{Tandon_2017} & 2nd FUV grating (\#66126) & 1st FUV grating (\#63771) & NUV grating (\#66125) \\
IDs in UVIT Pipeline & F4 & F6 & F4 \\
$m=-1$ peak $\lambda$ & -- &  -- & $2100\AA$ \\
$m=-2$ peak $\lambda$ & $\sim1400\AA$ & $\sim1500\AA$  & -- \\
Spectral resolution (FWHM) & $14.6\AA$  & $14.6\AA$ & $33\AA$ \\  \hline
\end{tabular}
\label{tab:grating_par}
\end{table*}

\section{Calibration observations and data reduction \label{sec:data}}
We used UVIT observations of NGC~40 and HZ~4, these observations  are as listed in Table~\ref{tab:obs}. The planetary nebula NGC~40 has a rich set of UV emission lines and is well suited for wavelength calibration. The white dwarf HZ~4 is a well-established spectrophotometric standard with nearly featureless spectrum, and is appropriate for flux calibration (e.g.,  \citealt{1990ApJS...73..413B}). HZ~4 has also been used for the photometric calibration of UVIT filters \citep{Tandon_2017,2020AJ....159..158T}

We used the Level1
data from the observations listed in Table~\ref{tab:obs}. 
We used the UVIT pipeline
CCDLAB \citep{2017PASP..129k5002P} to process the Level1 data. The pipeline performs corrections for field distortion, centroiding bias, flat field, pointing drift and accounts for frames rejected due to cosmic rays or missing from the level1 data.  We used the VIS images to generate  drift series which could then correct for the pointing drift in each orbit. We generated cleaned images for each orbit, aligned and merged them for each grating, as shown in Figure~\ref{fig:grating_images}. The x,y coordinates in these images represent 1/8 subpixel coordinates. Hereafter, we refer to these 1/8 subpixels simply as pixels which are $\approx 0.413\arcsec$ wide. The images show the undispersed zeroth order image together with the dispersed $\pm1$ and $\pm2$ order two dimensional spectra. The dispersion axis for the FUV-G1 and NUV-G is nearly horizontal, while it is roughly vertical in the case of FUV-G2. The grating images show maximum intensity in the blazed order ($m=-2$ for FUV gratings and $m=-1$ for the NUV grating). The FWHM of the intensity distribution along the cross-dispersion direction is a measure of combined PSF due to the telescope, detector and grating. Since gratings introduce some distortions, the PSF is slightly broader ($FWHM\sim 2\arcsec$) for grating images compared to that for the broadband filter images \citep{Tandon_2017}.

For the calibration and analysis of the UVIT data processed with the CCDLAB pipeline, we have developed a software package {\tt UVITTools} in the Julia language \citep{2012arXiv1209.5145B}.  We have used this package extensively here.  We also used the FTOOLS\footnote{\url{https://heasarc.gsfc.nasa.gov/lheasoft/}}, Sherpa \citep{2001SPIE.4477...76F} and XSPEC \citep{1996ASPC..101...17A} packages for generating response files and fitting the data.

\begin{table*}
\centering
\caption{UVIT/Grating observations of NGC~40 and HZ~4}
\begin{tabular}{cccccc} \hline
 Target & ObsID & Date of Observation & Grating  & Exposure Time (s) & window size \\  \hline
   
 NGC~40  & C02\_010T01\_900000 & 2016-12-07 &  FUV-G1 & 1194.0 & $350\times350$  \\
         &  & &  FUV-G2 &  1186.9 & $350\times350$ \\
         &   & &  NUV-G  & 2407.8 & $350\times350$ \\  \hline
 HZ~4  & T01\_054T01\_9000000 & 2016-02-02 & FUV-G1 &  1337.8 & $512\times512$  \\
       &  &  &  FUV-G2 &  1026.2 & $512\times512$ \\
       &  &  &  NUV-G  &  3082.9 & $512\times512$  \\  \hline
 
 \end{tabular}
 \label{tab:obs}
\end{table*}

\section{Extraction of 1d spectra \label{sec:extr_spec}}
 It is clear from Fig.~\ref{fig:grating_images} that the dispersion axes are not exactly aligned to either of the x or y axis. We measured the tilt angle  relative to the x-axis and found that the dispersion axes can be represented by the linear relations $y=mx +c$ where $m=\tan{\theta}$,  $c=y_o - mx_0$ and $x_0$ and $y_0$ are the pixel coordinates of the centroid of the zero order image.  We list the tilt angles in Table~\ref{tab:wave_calib}. The linear relations define the spectral trace i.e., the centroids of the cross-dispersion spatial profiles at each pixel on the dispersion axis. There is no additional significant distortion in the dispersion direction, so spatial profile fittings to trace  the dispersion direction is not required.

\begin{figure}
\centering
\includegraphics[width=8cm]{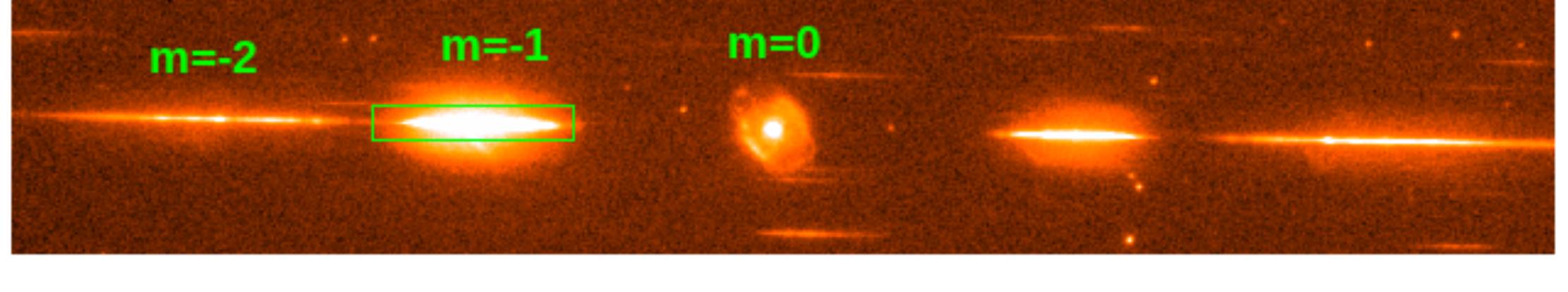}
\caption{NUV Grating image of NGC~40 with the grating orders marked. The extent along the dispersion direction and the width along the spatial direction used to extract the 1d spectrum is shown as the rectangular region covering the $m=-1$ grating order.}
\label{fig:1d_nuv_spec_extr}
\end{figure}

\begin{figure*}
\centering

\includegraphics[width=8cm]{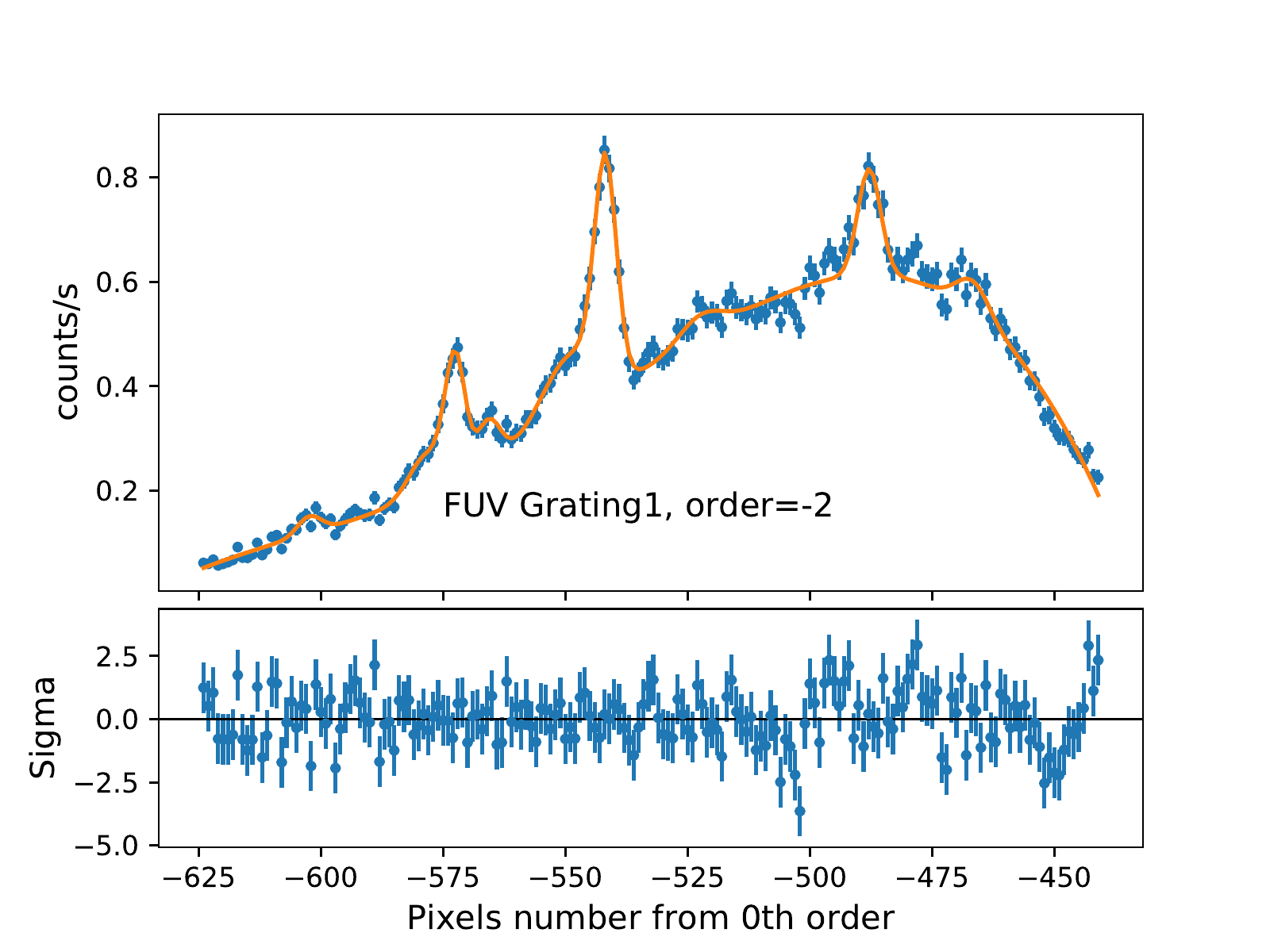}
\includegraphics[width=8cm]{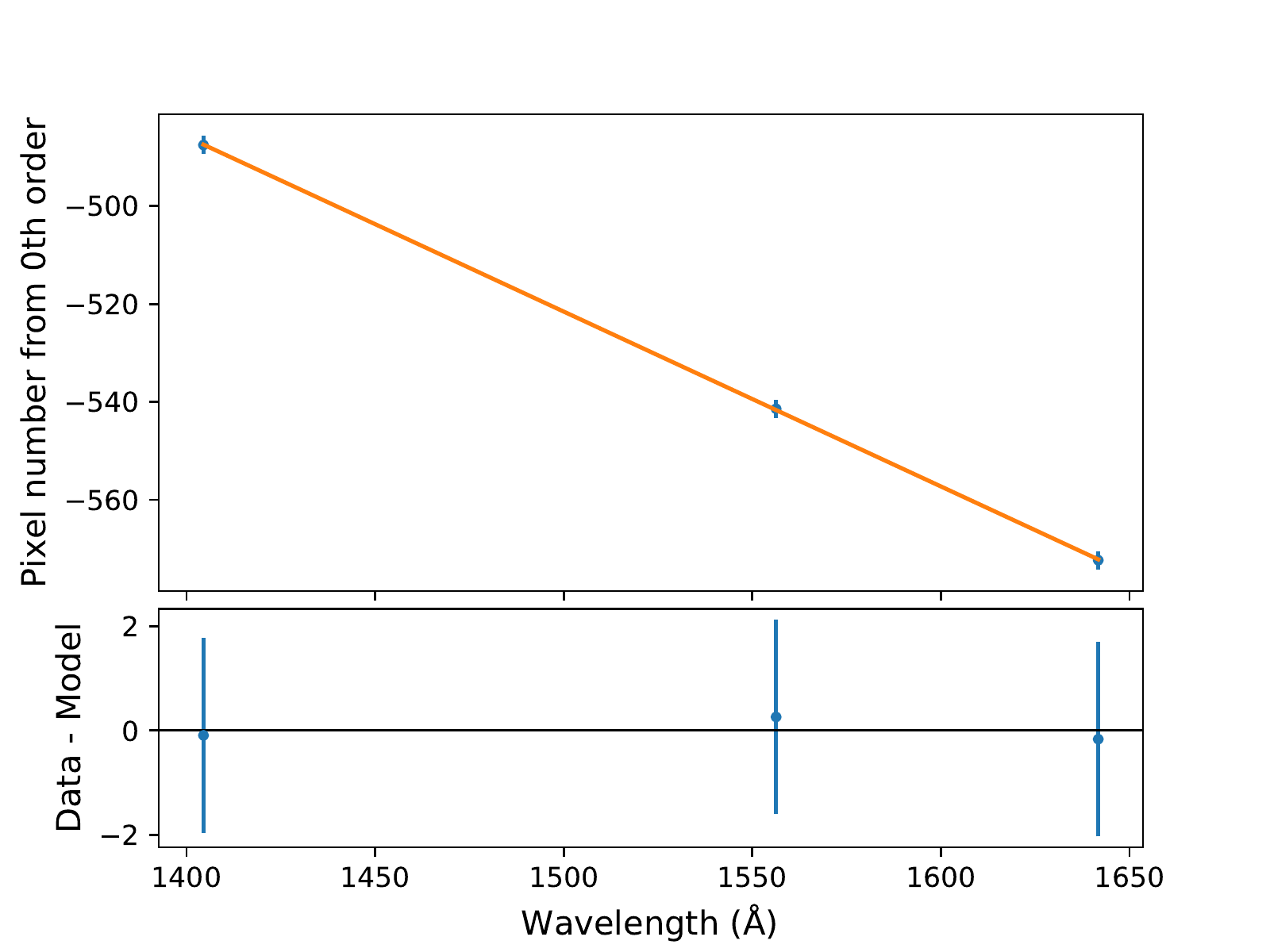}
\includegraphics[width=8cm]{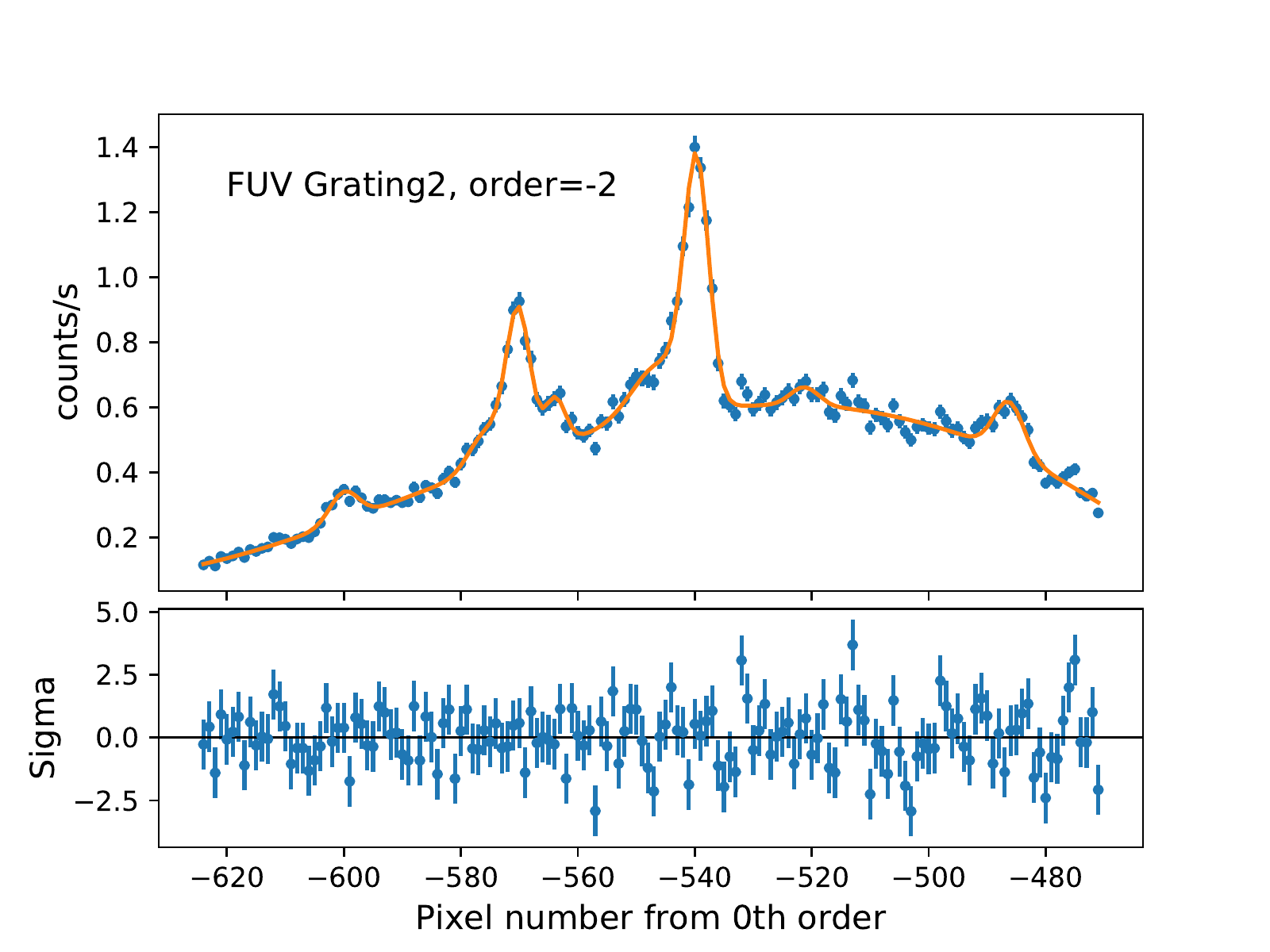}
\includegraphics[width=8cm]{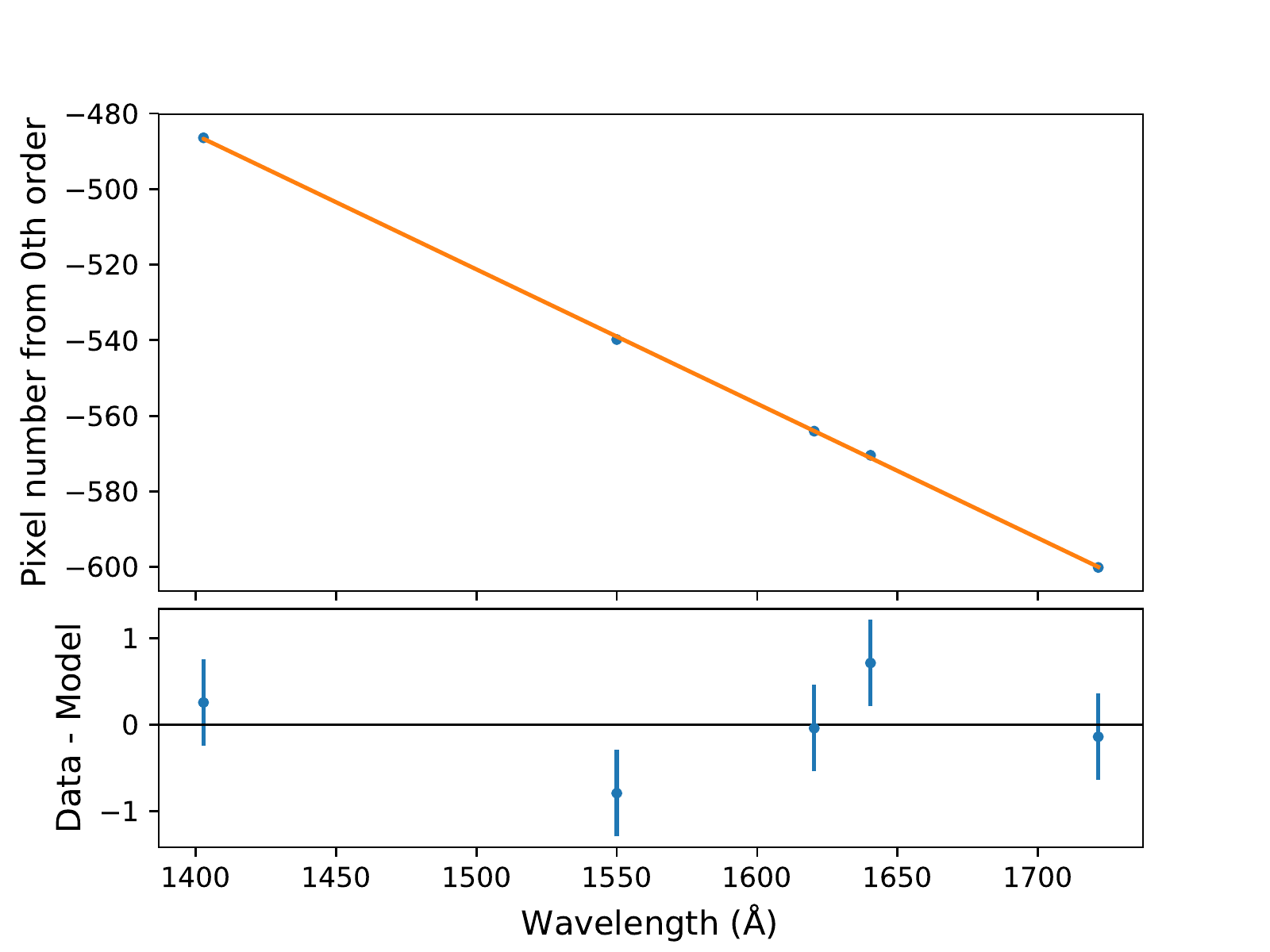}
\includegraphics[width=8cm]{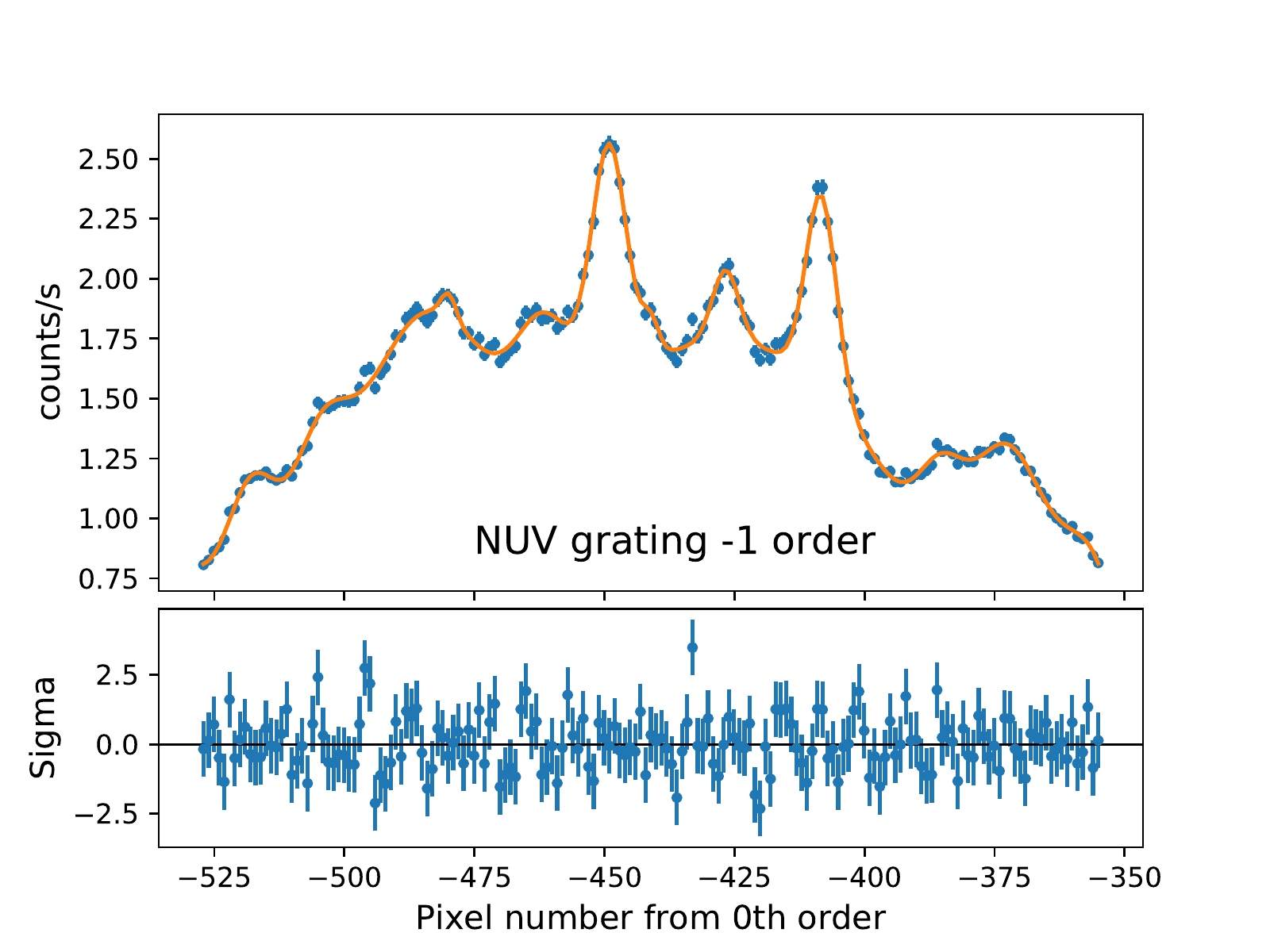}
\includegraphics[width=7.5cm]{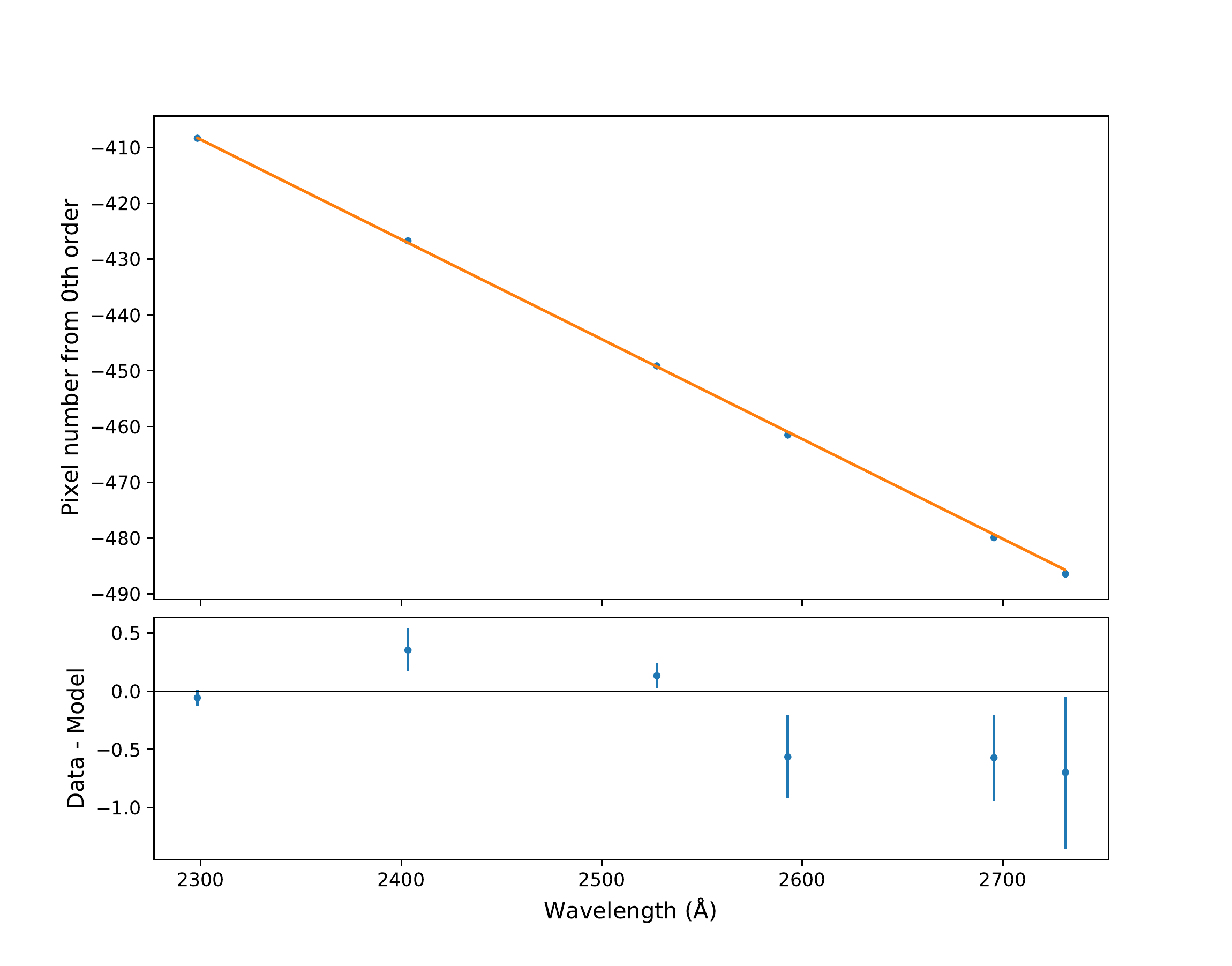}
\caption{The UVIT grating spectra of NGC~40 (left panels) and pixel-wavelength calibration (right panels) of the blazed orders (top panels: FUV-G1, $m=-2$; middle panels: FUV-G2, $m=-2$, bottom panels: NUV-G, $m=-1$). The 1d spectra i.e., count rate Vs pixel numbers from the centroid of the 0th order, are fitted with a number of Gaussian profiles for the emission lines and polynomial for the continua.}
\label{fig:wave_calib}
\end{figure*}

We define coordinates along the dispersion direction as the pixel coordinates relative to the zero order position.  We examined the grating images visually and determined the range of coordinates for different grating orders. These ranges are listed in Table~\ref{tab:wave_calib}. We used a 50~pixel width along the spatial direction centered on the trace defined by the linear relations and summed the counts to generate 1d spectra of a source of interest. Figure~\ref{fig:1d_nuv_spec_extr} shows the region used for the extraction of 1d spectrum of NGC~40 in the $m=-1$ order. We also extracted a 1d background spectrum from a source-free region in the image using the same relative coordinates in the dispersion direction and the same width along the spatial direction as used for the source. We subtracted the background counts from the source counts and propagated their errors on counts. The net 1d spectra of NGC~40 thus generated are shown in the left panels of Figure~\ref{fig:wave_calib}.

%
%

\begin{table*}
\centering
\caption{Parameters for 1d spectral extraction and the coefficients of the linear dispersion relation}
\begin{tabular}{cccccc} \hline
Grating & Tilt angle ($\theta$) & Order & Range in relative Pixel coordinate (X)  & \multicolumn{2}{c}{Linear dispersion relation}  \\ 
        &                        &      &   & Intercept ($a$)  & Slope ($b$) \\ \hline
FUV-G1 & $358.703^{\circ}$  &  $-2$ &  $-629$ to $-413$  & $43.4$ &  $-2.791$ \\
             & $358.703^{\circ}$  & $-1$ &  $-323$ to $-213$  & $-18.0$ & $-5.833$ \\  \\
FUV-G2 &  $267.531^{\circ}$ & $-2$ &  $-624$ to $-426$   & $31.2$ &  $-2.812$ \\
             &  $267.531^{\circ}$  & $-1$ & $-313$ to $-228$   & $45.0$  & $-5.625$ \\  \\
NUV-G  &  $358.904^{\circ}$  & $-1$ &  $-545$ to $-336$  & $45.1$ & $-5.523$ \\    \hline         
\end{tabular}
\label{tab:wave_calib}
\end{table*}

\section{Wavelength Calibration \label{sec:wave_calib}}
The raw 1d spectra shown in Figures~\ref{fig:wave_calib} are not in physical units. In order to convert the relative pixel coordinates along the dispersion direction, we used the emission lines from NGC~40. We measured the  emission line positions by fitting Gaussian profiles along with low order polynomial for the continuum. We then identified the strong emission lines in the UVIT spectra using the emission lines listed in \cite{1999ApJ...514..296F} based on {\it IUE} observations. We then fitted the following  linear relation between the line wavelengths and the relative pixel numbers. 
\begin{equation}
\lambda(\AA) = a + b  X,
\end{equation}
where $X$ is the pixel coordinate along the dispersion direction relative to the zero order position. 
The best-fit linear relations are shown in the right panels of Figures~\ref{fig:wave_calib} for the blazed orders of FUV and NUV gratings. The slope and intercept of the best-fitting linear dispersion relation are listed in Table~\ref{tab:wave_calib}. We converted the pixel coordinates in the raw grating spectra to wavelength in Angstroms using the above dispersion relations.


\begin{figure}
\centering
\includegraphics[width=8cm]{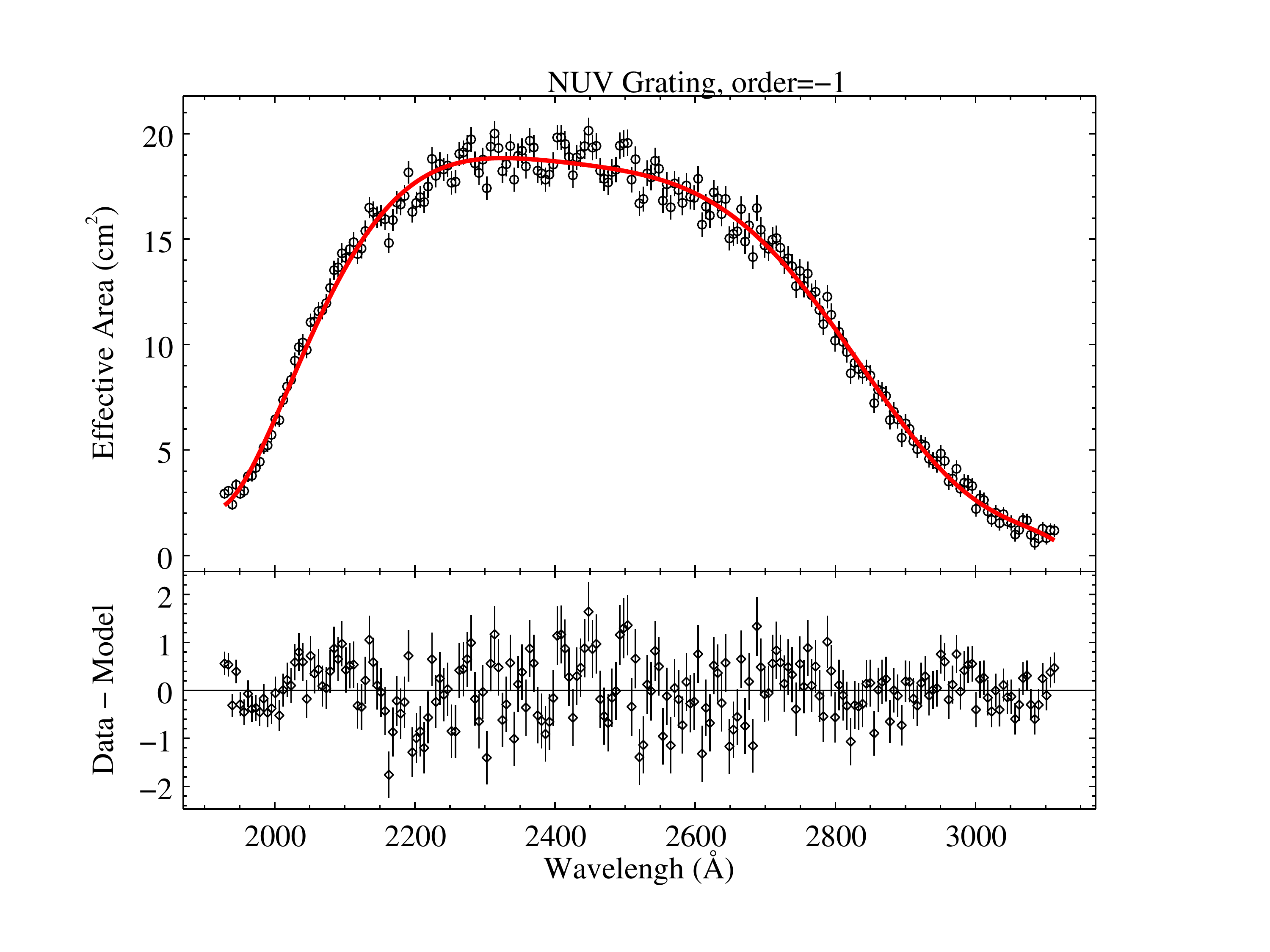}
\caption{The $m-=-1$ order NUV Grating effective area, the best-fitting polynomial and the residuals.}
\label{fig:nuv_effarea}
\end{figure}

\begin{figure*}
\centering
\includegraphics[width=8cm]{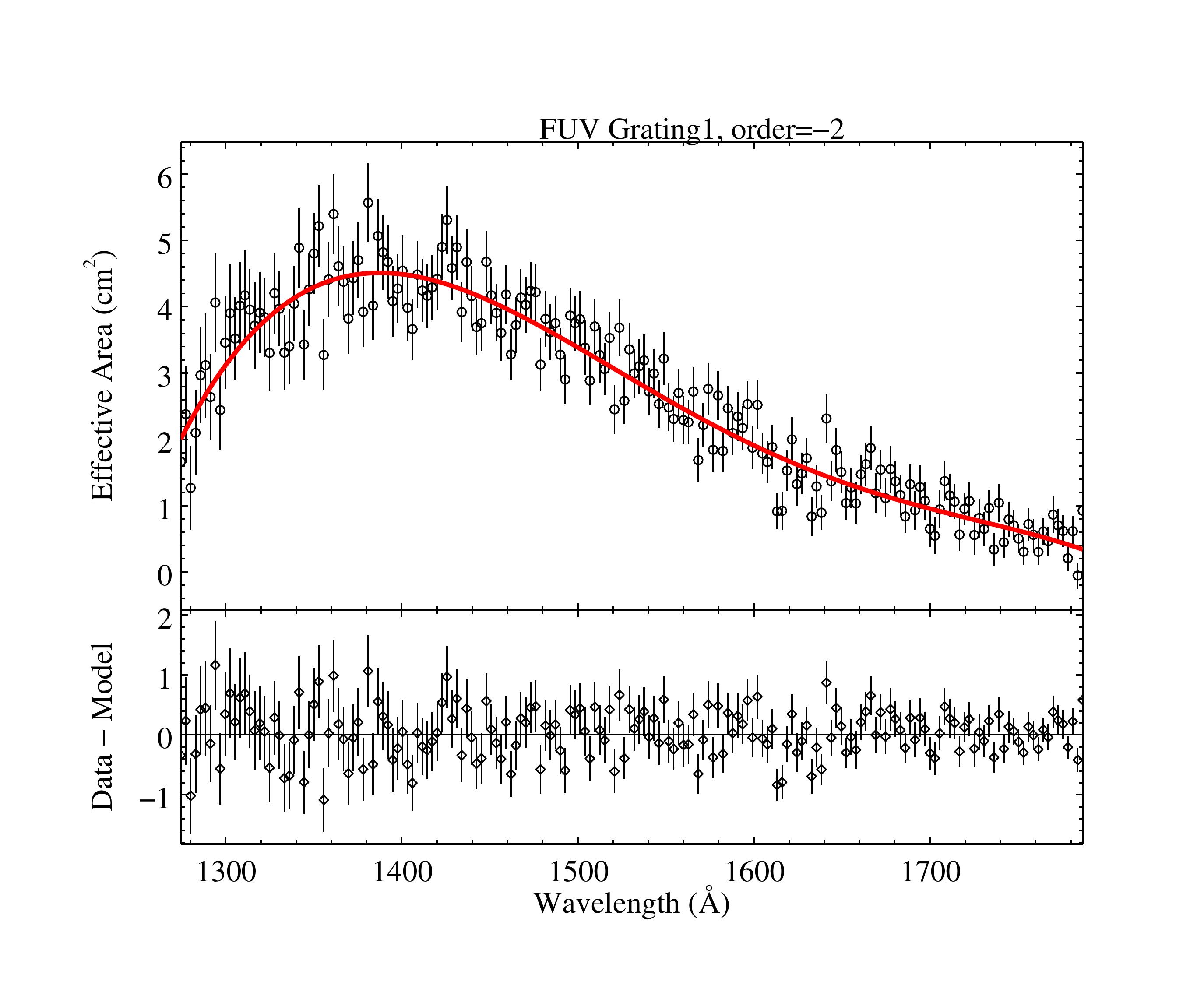}
\includegraphics[width=7.5cm]{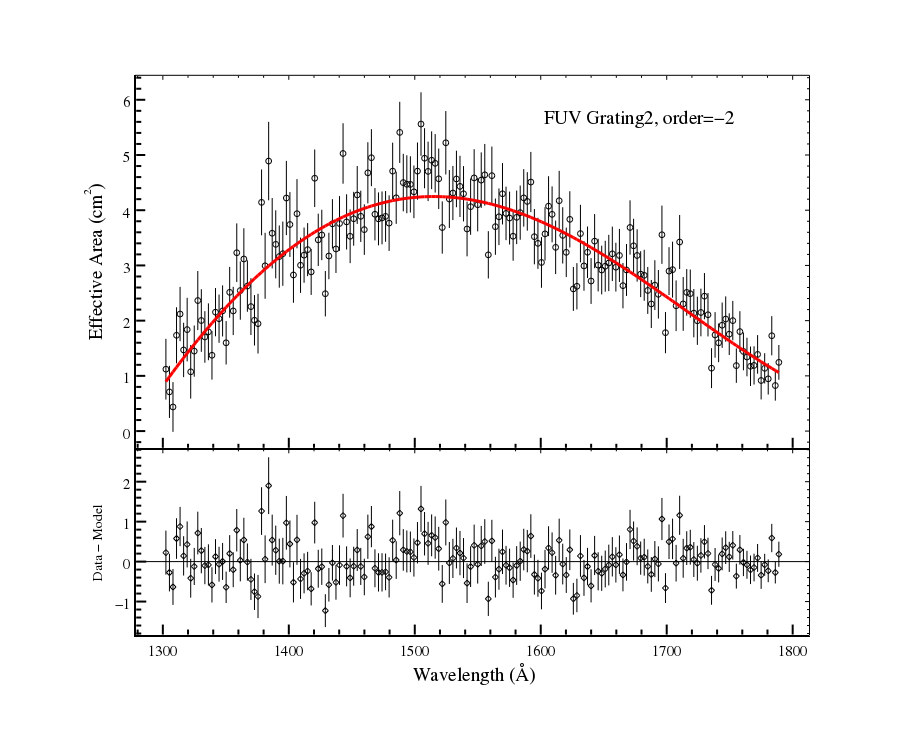}
\caption{The $m=-2$ order FUV Grating1 and Grating2 effective areas, the best-fitting polynomials and the residuals.}
\label{fig:fuv_effarea}
\end{figure*}

\begin{table*}
\centering
\caption{The coefficients of the best-fitting polynomials to the effective areas of UVIT gratings}  
\begin{tabular}{llcccc} \hline
Coefficient & \multicolumn{2}{c}{FUV-G1}  & \multicolumn{2}{c}{FUV-G2} & NUV-G \\
            & $m=-2$ & $m=-1$ & $m=-2$ & $m=-1$ & $m=-1$  \\ \hline
  $c_0$  & $53940.6148$   &   $-522747.0992$   & $15052.4039$  & $-1.2976$  &  $2.3671$ \\
$c_1$  &  $-212.0132$  &  $2424.3078$     & 	$-49.4344$  & $0.0418$   &  $-0.0362$ \\ 
$c_2$  &  $0.3446$ & -$4.8004$    &  $0.0645$  &  $-0.00036$ & $0.0011$ \\
$c_3$  &  $-0.0003$ & $0.0053$  & $-4.1806\times10^{-5}$  &  $1.3620\times10^{-6}$  & 	$-5.1848\times10^{-6}$ \\
$c_4$  &  $1.4266\times10^{-7}$ & $-3.4478\times10^{-6}$  & $1.3476\times10^{-8}$  &  $-2.2112\times10^{-9}$  &  	$1.08550\times10^{-8}$ \\ 
$c_5$  &  $-3.6369\times10^{-11}$ & $1.3509\times10^{-9}$  &  $-1.7298\times10^{-12}$  &  $1.2713\times10{-12}$  &  $-1.19142\times10^{-11}$ \\
$c_6$  &   $3.8411\times10^{-15}$ & $-2.9306\times10^{-13}$  &  --  &     -- &   $6.5940\times10^{-15}$  \\
$c_7$  &   --                     & $2.7159\times10^{-17}$   &  --   &   --  &     $-1.4486\times10^{-18}$ \\
$\lambda_{0}$   &       0.0      &        0.0        &    0.0     & $1250.0$  &  $1900.0$  \\  \hline
\end{tabular}
\label{tab:ea_poly}
\end{table*}

\section{Flux Calibration and Effective Area curves \label{sec:flux_calib}}
Flux calibration is the process of converting the observed count rate  to flux density as a function of wavelength. The background-subtracted net count rate is related to the incident flux density $f_{\lambda}$[${\rm ergs~cm^{-2}~s^{-1}~\AA^{-1}}$] as 

\begin{equation}
C(X) = \int{R_{X{\lambda}} A_{\lambda} f_{\lambda}\left(\frac{\lambda}{hc}\right)}d\lambda
\label{eqn:spec_resp}
\end{equation}
where $C(X)$ is the net count rate at $X$ along the dispersion direction, $R_{{\lambda}X}$ gives the probability that a UV photon of wavelength $\lambda$ gets detected at $X$, $A_\lambda$ is effective area of the spectrometer accounting for the collecting area of the telescope, reflectivity of the mirrors, grating efficiency, detector quantum efficiency and attenuation efficiency of any other optical element. Thus $A_{\lambda}$ is telescope area corrected for all losses. In principle, $A_{\lambda}$ can be determined from pre-launch lab measurements, however, the various efficiencies generally change with time during and after launch. Hence, $A_\lambda$ is generally determined or corrected based on observations of standard sources with well measured fluxes. Eqn.~\ref{eqn:spec_resp} is appropriate for the photon counting data from UVIT, and can be written in  matrix form with $\lambda$, $X$ now representing wavelength and spectral bins,

\begin{equation}
C_X  = \sum_{\lambda}{ R_{X\lambda} A_{\lambda} f_{\lambda} \frac{\lambda}{hc}}
\end{equation}

In the case of dispersive spectrometers such as the FUV/NUV gratings with suppression of order-overlap, the response matrix can be assumed to be diagonal. Hence $R_{X\lambda}=1$ if $\lambda$ and X follow the dispersion relation. Thus,
\begin{equation}
  f_{\lambda}=\frac{C_{\lambda}(hc/\lambda)}{A_{\lambda}},
  \label{eqn:effarea}
\end{equation}
where $C_{\lambda}$ is the count rate spectrum. 

To determine the effective areas of the gratings, we used the UVIT grating observations of spectrophotometric standard  HZ4 (whose UV spectral flux values were obtained from spectrum file hz4\_stis 005.fits available at  HST-CALSPEC\footnote{\url{http://www.stsci.edu/hst/observatory/cdbs/calspec.html}}). We matched the wavelengths bins of $C_{\lambda}$ measured with UVIT gratings and the spectrum by linearly interpolating the standard spectrum. One can then calculate the effective area of the UVIT-gratings using Eqn.~\ref{eqn:effarea}. However, we note that HZ~4 shows strong absorption lines at $1216\AA$ (Ly$\alpha$) and $1400\AA$, and the spectral resolution of the standard spectrum measured with {\it IUE} in the far UV band is superior compared to that of the UVIT grating spectra. We therefore smoothed the HZ4 standard spectrum to match the UVIT grating spectral resolution.
We used Gaussian kernels with widths that resulted in smooth effective areas using Eqn.~\ref{eqn:effarea} for different grating orders. The effective areas thus derived are shown in Figure~\ref{fig:nuv_effarea} for the $-1$ order of NUV-G, and in Figure~\ref{fig:fuv_effarea} for the $-2$ orders of FUV-G1 and FUV-G2.

We have fitted the UVIT-grating effective areas with low order polynomials of the form $A_{\lambda}=\sum{c_n (\lambda-\lambda_0)^n}$  so that the observed count rate spectrum can be converted to the fluxed-spectrum by dividing the best-fitting polynomials using Eqn.~\ref{eqn:effarea}. The best-fitting coefficients are listed in 
Table~\ref{tab:ea_poly}, and the best-fitting polynomials are shown in  Fig.~\ref{fig:nuv_effarea}
and Fig.~\ref{fig:fuv_effarea}.

\begin{figure*}
\centering
\includegraphics[width=19cm]{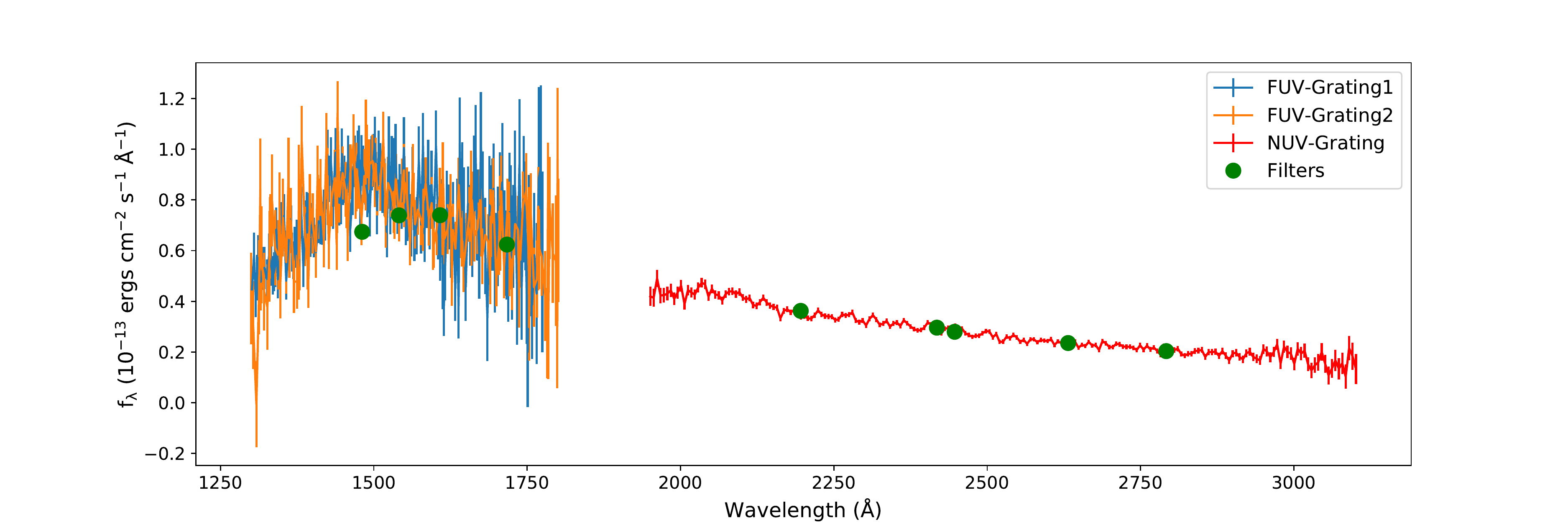}
\caption{A comparison of flux densities at various wavelengths derived from the FUV/NUV broadband filters and grating observations of HZ~4.}
\label{fig:cross_calib}
\end{figure*}

\section{Cross-calibration between gratings and broadband filters \label{sec:cross_calib}}
The flux densities measured by broadband filters at their mean wavelengths must be the same as those measured by the gratings at the same wavelengths if the mean wavelengths represent the effective wavelengths of the filters. Small discrepencies may be expected as the mean wavelengths of the UVIT filters are not defined to filfil the above condition (private communication with S. N. Tandon). Nevertheless it is useful to compare the flux measured with the gratings and broadband filters.
As mentioned earlier, gratings introduce distortion making the PSF slightly poorer. Hence, the same sizes for the extraction regions i.e.,  the spatial width along the cross-dispersion direction in the case of grating observations and the diameter of the circular region in the case of broadband filter observations, may not provide the same flux densities. Therefore it is important to cross-calibrate the gratings and the broadband filters. 


For this purpose, we used the calibration information and count rates already derived by \cite{2020AJ....159..158T}. For HZ~4, we used their corrected count rates that were derived after applying corrections for flat field and saturation effects. We converted these count rates to flux densities using the  unit conversion (UC) factors which we calculated from the ZP magnitudes listed in Table~3 of \cite{2020AJ....159..158T}. We derived the UC factors from the ZP magnitude as follows \citep{Tandon_2017},
\begin{equation}
UC=  10^{-0.4\frac{(ZP + 2.407)}{\lambda_{mean}^2}},
\label{eqn:uc}
\end{equation}
where UC is in ${\rm ergs~cm^{-2}~s^{-1}~\AA^{-1}}$ and $\lambda_{mean}$ is in $\AA$. The flux density $f_{\lambda}$ in ${\rm ergs~cm^{-2}~s^{-1}~\AA^{-1}}$ is
\begin{equation}
f_{\lambda} = CPS \times UC,
\label{eqn:cps2flux}
\end{equation}
where CPS is the count rate in counts~s$^{-1}$. In Table~\ref{tab:uc}, we list the net count rate from \citep{2020AJ....159..158T}, the UC factors and flux density calculated using Eqn.~\ref{eqn:uc} and Eqn.~\ref{eqn:cps2flux} for HZ~4. 

In order to check possible calibration differences between the  flux densities derived from the gratings and broadband filters, we plot FUV/NUV grating spectra and flux densities derived for different filters in Figure~\ref{fig:cross_calib}. It is clear that, except for the F148W (CaF2-1) filter, the flux measurements agree very well between the gratings and the broadband filters. The F148W flux density is $\sim 25\%$ lower compared to that measured with the FUV gratings. The discrepancy is at a level of $2.4\sigma$. In comparison to the standard spectrum  ({\tt hz4\_stis\_005.fits)} used for the calibration of gratings, the F148W flux density is also lower by $\sim21\%$. This apparent discrepancy is most likely due to the presence of sharp spectral features in the standard spectrum near the mean wavelength of the F148W filter. The spectral features are not well resolved by the FUV gratings. To further check, we calculated the expected count rates of HZ~4 using Eqn.~\ref{eqn:spec_resp} based on the effective areas from \citep{2020AJ....159..158T} and the FUV-G1 and NUV-G spectra. The predicted count rates are listed in Table~\ref{tab:uc}. It is clear that the observed and predicted count rates agree very well. This shows the importance of using effective areas as discussed below.
Based on the above analysis, we recommend a cross-dispersion width of $50$ pixels for spectral extraction of point sources. In case of poor tracking correction and extended sources, a larger cross-dispersion width should be used.

\begin{table*}
\caption{The unit conversion factor (UC) for UVIT filters, and the observed count rate, flux density and predicted count rate for HZ~4 in different filters. }
\begin{tabular}{ccccccc}  \hline
\multicolumn{2}{c}{Filter}  & $\lambda_{mean}$   & UC                                   &   \multicolumn{3}{c}{HZ~4}  \\ 
      &                    &  ($\AA$)            &  ($\rm ergs~cm^{-2}~s^{-1}~\AA^{-1}$) & Obs. CPS$~^a$  & $f_\lambda$$~^b$ & Pred. CPS$~^c$  \\ \hline
F148W	& CaF2-1	& 1481	& ($2.866\pm0.026$)$\times10^{-15} $ & 23.5	    & $6.741\pm0.062$  & $21.5\pm1.3$ \\
F154W	& BaF2	    & 1541	& ($3.574\pm0.033$)$\times10^{-15} $ & 20.7	     & $7.392\pm0.068$ & $19.1\pm1.4$ \\
F169M	& Sapphire	& 1608	& ($4.57\pm0.0427$)$\times10^{-15} $ & 16.2	    & $7.397\pm0.068$ &  $15.5\pm1.2$ \\
F172M	& Silica	& 1717	& ($1.143\pm0.021$)$\times10^{-14}$ & 5.5	    & $6.24\pm0.11$  & $5.7\pm0.8$ \\
N242W	& Silica-1	& 2418	& ($2.3179\pm0.0043$)$\times10^{-16}$ & 127.8   & $2.9622\pm0.0055$ & $131.6\pm1.2$ \\
N219M	& NUVB15	& 2196	& ($4.924\pm0.091$)$\times10^{-15}$ & 7.4	    &  $3.624\pm0.067$ & $7.5\pm0.1$ \\
N245M	& NUVB13	& 2447	& ($7.571\pm0.035$)$\times10^{-16}$ & 37.0	   & $2.799\pm0.013$ & $37.8\pm0.3$ \\
N263M	& NUVB4	    & 2632	& ($8.674\pm0.080$)$\times10^{-16}$ & 27.2	    &  $2.356\pm0.022$ &  $27.6\pm0.4$ \\
N279N	& NUVN2	    & 2792	& ($3.793\pm0.035$)$\times10^{-15}$ & 5.4	   & $2.037\pm0.019$  & $5.5\pm0.1$ \\ \hline
\end{tabular}

$^a$ Observed count rate in counts~s$^{-1}$. Errors on the observed CPS are less than $2\%$.\\
$^b$ In unit of $10^{-14}\rm ergs~cm^{-2}~s^{-1}~\AA^{-1}$. \\
$^c$ Predicted count rate using the effective areas provided in \citep{2020AJ....159..158T} and the FUV-G1 and NUV-G spectra of HZ~4 shown in Fig.~\ref{fig:cross_calib}.

\label{tab:uc}
\end{table*}




\section{Grating spectral responses and multi-wavelength spectroscopy \label{sec:spec_resp}}
Working with  data from different instruments onboard \astrosat{} require tools and techniques to faciliate joint analysis of multi-wavelength data. In particular, the broadband spectral coverage of \astrosat{}, from near UV to hard X-rays, requires tools for simultaneous fitting of spectral models to the multi-wavelength data. Due to the complex interactions when X-rays go through the detector material, the response function of X-ray detectors are generally complex, and the X-ray spectral data cannot be directly converted to the source spectrum. X-ray spectral analysis begins with an assumed source spectral model which is then folded with the instrument response that results in model spectral data which is then compared with the observed spectral data. The source spectrum is thus inferred from the best-fitting model. Grating spectrometers, such as the UVIT gratings, have much simpler response functions, and one generates a fluxed spectrum directly from the observations using the dispersion relation and the wavelength-dependent effective area or sensitivity curve as described in the previous section. The source spectral properties are inferred from the fluxed spectrum after correcting for the instrumental spectral resolution. It is possible to treat the photon counting data from UVIT similar to the X-ray data that are also photon counting by generating appropriate spectral responses in the form of a redistribution matrix (RMF) and ancillary response file (ARF), and use  Eqn.~\ref{eqn:resp} below which is similar to Eqn.~\ref{eqn:spec_resp} to infer the source spectrum. One can then perform joint spectral analysis of UVIT and X-ray data from SXT, LAXPC and CZTI. In the following, we generate the count spectrum (i.e., distribution of photon counts in different spectral channels), redistribution matrix and ancillary response for UVIT gratings and broadband filters. Such an approach is more accurate than directly converting the count rates to flux densities with unit conversion factors and zero point magnitudes  for various filters as the latter quantities depend on the spectral shape.  The unit conversion factors and zero point magnitudes are derived for the particular spectral shape of the spectrophotometric standard HZ~4 and are unlikely to be strictly correct for objects with different spectral shape.

\begin{figure*}
\centering
\includegraphics[width=8.5cm]{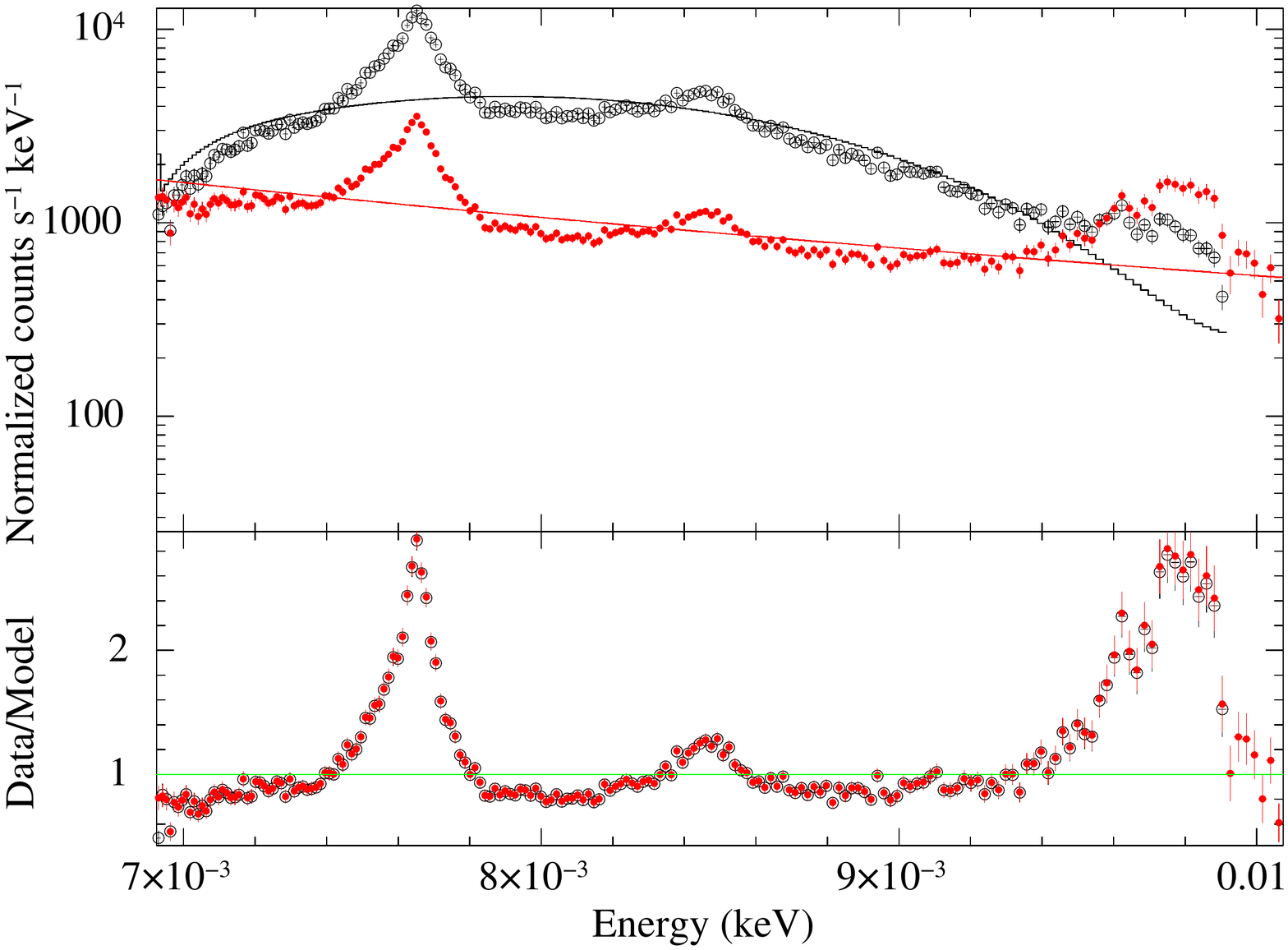}
\includegraphics[width=8.5cm]{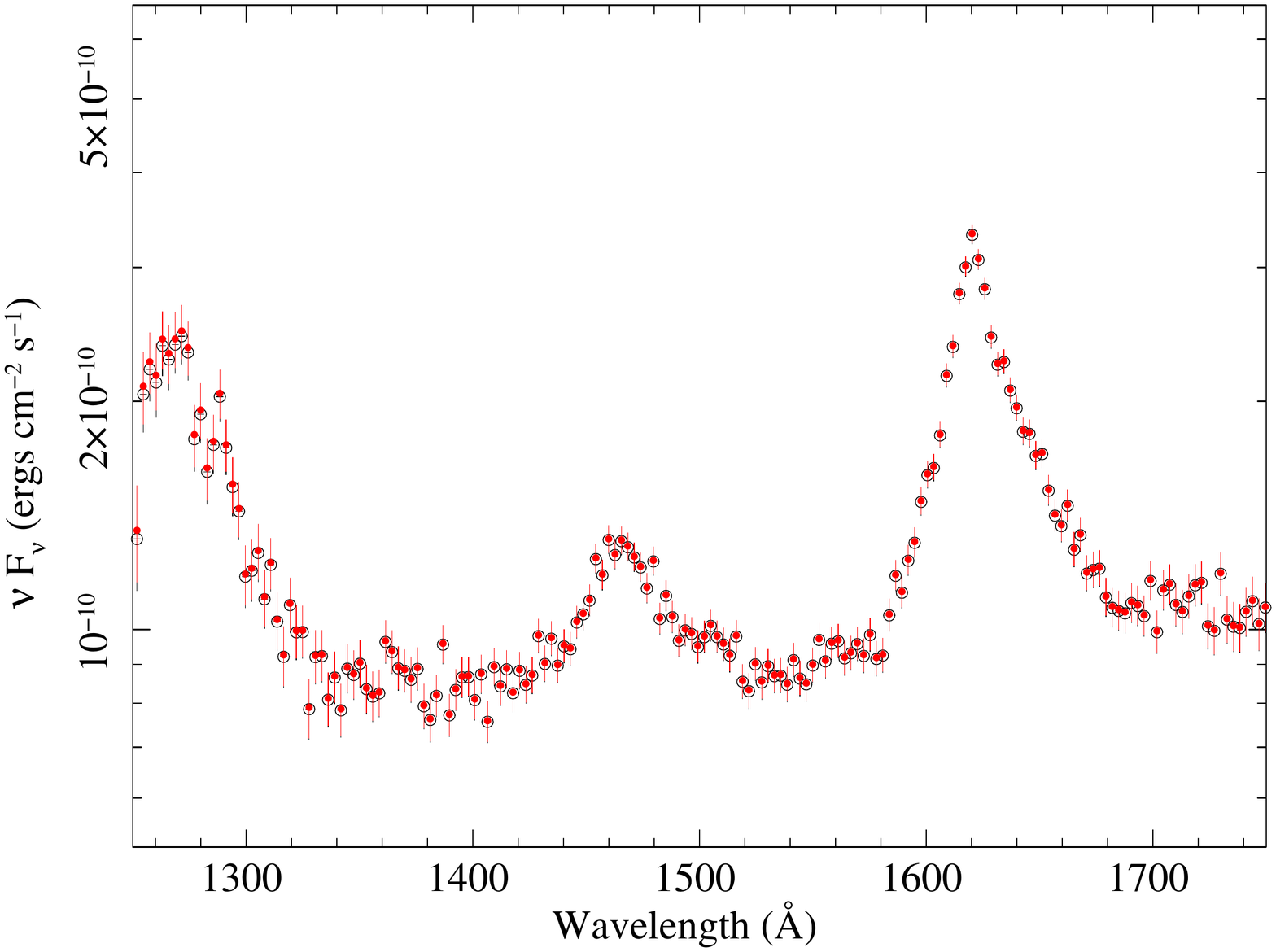}
\caption{Comparison of count PHA spectrum of Fairall~9 with instrument response and fluxed  PHA spectrum with diagonal response. {\it Left panels:} Count PHA data (black open circles), flux PHA data (red filled circles), the same powerlaw fitted to the both datasets and the data-to-model ratios. {\it Right panel:} Unfolded spectra using a powerlaw of photon index, $\Gamma=0$ ($N_E = AE^{-\Gamma}$) and unit norm at 1~keV. Both the count PHA and fluxed spectra were extracted from the $-2$ order of FUV-G2 image.}
\label{fig:f9_spec}
\end{figure*}

For the UVIT gratings, we constructed a count spectrum (counts vs spectral channels) from the uncalibrated 1d spectrum (counts vs relative pixel coordinates) in the same format as the X-ray spectral PHA files. For this purpose, we added a positive integer number to the relative pixel coordinate and converted to spectral channels (I) which start from 1. This count spectrum is related to the source spectrum in a way similar to Eqn.~\ref{eqn:spec_resp},

\begin{equation}
D(I) = T\int{R(I,E) A(E) N_E dE} + B(I),
\label{eqn:resp}
\end{equation}
where $D(I)$ is the source + background count in spectral channel $I$, $B(I)$ is the background counts in channel $I$, $T$ is exposure time. $R(I,E)$ is the redistribution matrix and $A(E)$ is effective area at energy $E$, $N_E$ is the photon spectrum of the source.

The redistribution matrix $R(I,E)$ represents the spectral response which is Gaussian with an FWHM that is the same as the spectral resolution of the grating spectrometers. To measure the spectral resolution of UVIT gratings, we fitted Gaussian profiles to the emission lines observed from NGC~40 and derived the FWHM$=14.63\AA$ (FUV gratings, $m=-2$), $33\AA$ (NUV-G, $m=-1$). In order to generate the response matrix for the NUV-G, we divided the wavelength range in 2341 energy bins. For each energy bin, we generated the Gaussian response in 210 channels which were defined based on the dispersion solution. We also multiplied the effective area with the redistribution matrix as in Eqn.~\ref{eqn:resp} and created the grating response matrix which is the product $R(I,E)A(E)$. The responses thus created for each calibrated grating order are compatible with the X-ray spectral fitting packages such as XSPEC \citep{1996ASPC..101...17A}, Sherpa \citep{2001SPIE.4477...76F} and ISIS \citep{2000ASPC..216..591H}. The source and background PHA files along with the spectral response can directly be used in one of the above spectral fitting packages. This is helpful for joint UV/X-ray spectral modeling. We also created single channel response matrices for the broadband filters using the updated effective areas provided in \cite{2020AJ....159..158T}. These response files along with the grating or filter spectral PHA data in the fits format give complete flexibility to treat and analyze the photon counting UVIT data in a way similar to the X-ray photon counting data.

Another way to use the UV grating spectra and photometric flux in the joint UV/X-ray spectral analysis is to convert the fluxed spectra in PHA format and generate diagonal responses using the FTOOLS package. However, this has the disadvantage of not considering the actual spectral response of the instrument. Hence, the uncertainty associated with the UC factor  for the particular spectral shape of the standard star for a given broadband filter will enter into the spectral fitting. In the case of grating spectra, the derived spectral line widths will not be free of instrumental resolution as the instrument spectral response is not being used. 

We demonstrate the equivalence of these two approaches as follows. For this purpose we used UVIT grating observations of a Seyfert 1 AGN Fairall~9 (ObsID: G06\_157T01\_9000000). We processed the Level1 data in the same way as in the case of NGC~40 and HZ~4. We generated an  $m=-2$ order FUV-G2 spectrum of Fairall~9 in the PHA format. We also generated the associate background spectrum from the source-free regions as described earlier, and updated the PHA header of the source spectral file to include the grating response and the background spectral file. We loaded these spectral data into XSPEC and plotted the spectral data as shown in Figure~\ref{fig:f9_spec}. We also generated a fluxed spectrum  i.e., $f_\lambda$ Vs $\lambda$ of Fairall~9 for the FUV-G2 in the $m=-2$ order and converted these data to PHA file and diagonal response matrix using FTOOLS. The two spectra of Fairall~9 are compared in Fig.~\ref{fig:f9_spec}. It is clear that the two approaches agree well.

\section{Summary \& Conclusion \label{sec:summary}}
We calibrated the two FUV gratings in the orders $-2$ and $-1$ and one NUV grating in the $-1$ order. We derived dispersion solutions and effective areas for the grating orders which can be used for spectral calibration of any source observed with the UVIT gratings. We also checked the cross-calibration between the gratings and broadband filters and found excellent agreement in flux measurements for all broadband filters except the FUV filter F148W. We provide the updated UC and ZP for this filter. We also generated the spectral response files for the gratings and the broadband filters that can be directly used for spectral analysis using XSPEC, Sherpa or ISIS. A software package {\tt UVITTools} for the analysis of UVIT data processed with the CCDLAB pipeline has been developed.







\section*{Acknowledgments}
The author is grateful to Shyam Tandon for numerous discussions on various aspects of UVIT, and allowing to use the calibration data. The author is thankful to Phil Charles and Shyam Tandon for their suggestions on the submitted version of the manuscript. This publication uses the data from the AstroSat mission of the Indian Space Research Organisation (ISRO), archived at the Indian Space Science Data Centre (ISSDC). This publication uses UVIT data processed by the payload operations centre at IIA. The UVIT is built in collaboration between IIA, IUCAA, TIFR, ISRO and CSA.

\vspace{-1em}










\begin{thebibliography}{}
\expandafter\ifx\csname natexlab\endcsname\relax\def\natexlab#1{#1}\fi

\bibitem[{{Agrawal}(2006)}]{2006AdSpR..38.2989A}
{Agrawal}, P.~C. 2006, Advances in Space Research, 38, 2989

\bibitem[{{Arnaud}(1996)}]{1996ASPC..101...17A}
{Arnaud}, K.~A. 1996, in Astronomical Society of the Pacific Conference Series,
  Vol. 101, Astronomical Data Analysis Software and Systems V, ed. G.~H.
  {Jacoby} \& J.~{Barnes}, 17

\bibitem[{{Bezanson} {$et~al$.}(2012){Bezanson}, {Karpinski}, {Shah}, \&
  {Edelman}}]{2012arXiv1209.5145B}
{Bezanson}, J., {Karpinski}, S., {Shah}, V.~B., \& {Edelman}, A. 2012, arXiv
  e-prints, arXiv:1209.5145

\bibitem[{{Bohlin} {$et~al$.}(1990){Bohlin}, {Harris}, {Holm}, \&
  {Gry}}]{1990ApJS...73..413B}
{Bohlin}, R.~C., {Harris}, A.~W., {Holm}, A.~V., \& {Gry}, C. 1990, \apjs, 73,
  413

\bibitem[{{Feibelman}(1999)}]{1999ApJ...514..296F}
{Feibelman}, W.~A. 1999, \apj, 514, 296

\bibitem[{{Freeman} {$et~al$.}(2001){Freeman}, {Doe}, \&
  {Siemiginowska}}]{2001SPIE.4477...76F}
{Freeman}, P., {Doe}, S., \& {Siemiginowska}, A. 2001, in Society of
  Photo-Optical Instrumentation Engineers (SPIE) Conference Series, Vol. 4477,
  Astronomical Data Analysis, ed. J.-L. {Starck} \& F.~D. {Murtagh}, 76--87

\bibitem[{{Houck} \& {Denicola}(2000)}]{2000ASPC..216..591H}
{Houck}, J.~C., \& {Denicola}, L.~A. 2000, in Astronomical Society of the
  Pacific Conference Series, Vol. 216, Astronomical Data Analysis Software and
  Systems IX, ed. N.~{Manset}, C.~{Veillet}, \& D.~{Crabtree}, 591

\bibitem[{{Postma} \& {Leahy}(2017)}]{2017PASP..129k5002P}
{Postma}, J.~E., \& {Leahy}, D. 2017, \pasp, 129, 115002

\bibitem[{{Singh} {$et~al$.}(2014){Singh}, {Tandon}, {Agrawal}, {Antia},
  {Manchanda}, {Yadav}, {Seetha}, {Ramadevi}, {Rao}, {Bhattacharya}, {Paul},
  {Sreekumar}, {Bhattacharyya}, {Stewart}, {Hutchings}, {Annapurni}, {Ghosh},
  {Murthy}, {Pati}, {Rao}, {Stalin}, {Girish}, {Sankarasubramanian},
  {Vadawale}, {Bhalerao}, {Dewangan}, {Dedhia}, {Hingar}, {Katoch}, {Kothare},
  {Mirza}, {Mukerjee}, {Shah}, {Shah}, {Mohan}, {Sangal}, {Nagabhusana},
  {Sriram}, {Malkar}, {Sreekumar}, {Abbey}, {Hansford}, {Beardmore}, {Sharma},
  {Murthy}, {Kulkarni}, {Meena}, {Babu}, \& {Postma}}]{2014SPIE.9144E..1SS}
{Singh}, K.~P., {Tandon}, S.~N., {Agrawal}, P.~C., {$et~al$.} 2014, in Society
  of Photo-Optical Instrumentation Engineers (SPIE) Conference Series, Vol.
  9144, \procspie, 91441S

\bibitem[{{Subramaniam} {$et~al$.}(2016){Subramaniam}, {Tandon}, {Hutchings},
  {Ghosh}, {George}, {Girish}, {Kamath}, {Kathiravan}, {Kumar}, {Lancelot},
  {Mahesh}, {Mohan}, {Murthy}, {Nagabhushana}, {Pati}, {Postma}, {Rao},
  {Sankarasubramanian}, {Sreekumar}, {Sriram}, {Stalin}, {Sutaria}, {Sreedhar},
  {Barve}, {Mondal}, \& {Sahu}}]{2016SPIE.9905E..1FS}
{Subramaniam}, A., {Tandon}, S.~N., {Hutchings}, J., {$et~al$.} 2016, in
  Society of Photo-Optical Instrumentation Engineers (SPIE) Conference Series,
  Vol. 9905, Space Telescopes and Instrumentation 2016: Ultraviolet to Gamma
  Ray, ed. J.-W.~A. {den Herder}, T.~{Takahashi}, \& M.~{Bautz}, 99051F

\bibitem[{Tandon {$et~al$.}(2017)Tandon, Subramaniam, Girish, Postma,
  Sankarasubramanian, Sriram, Stalin, Mondal, Sahu, Joseph, Hutchings, Ghosh,
  Barve, George, Kamath, Kathiravan, Kumar, Lancelot, Leahy, Mahesh, Mohan,
  Nagabhushana, Pati, Rao, Sreedhar, \& Sreekumar}]{Tandon_2017}
Tandon, S.~N., Subramaniam, A., Girish, V., {$et~al$.} 2017, The Astronomical
  Journal, 154, 128

\bibitem[{{Tandon} {$et~al$.}(2020){Tandon}, {Postma}, {Joseph}, {Devaraj},
  {Subramaniam}, {Barve}, {George}, {Ghosh}, {Girish}, {Hutchings}, {Kamath},
  {Kathiravan}, {Kumar}, {Lancelot}, {Leahy}, {Mahesh}, {Mohan},
  {Nagabhushana}, {Pati}, {Rao}, {Sankarasubramanian}, {Sriram}, \&
  {Stalin}}]{2020AJ....159..158T}
{Tandon}, S.~N., {Postma}, J., {Joseph}, P., {$et~al$.} 2020, \aj, 159, 158

\end{thebibliography}

\end{document}